\documentclass[conference]{IEEEtran}
\IEEEoverridecommandlockouts
\usepackage{cite}
\usepackage{amsmath,amssymb,amsfonts}
\usepackage{graphicx}
\usepackage{textcomp}
\usepackage{xcolor}

\usepackage{algorithm}
\usepackage{algorithmicx}
\usepackage{algpseudocode}
\usepackage{multirow}
\usepackage{subfig}

\usepackage{url}

\usepackage{tikz}
\usetikzlibrary{arrows,calc,positioning,shadows,shapes}
\usepackage{pgfplots}
\pgfplotsset{compat=1.6}
\usepackage{balance}

\newtheorem{definition}{Definition}

\def\BibTeX{{\rm B\kern-.05em{\sc i\kern-.025em b}\kern-.08em
    T\kern-.1667em\lower.75ex\hbox{E}\kern-.125emX}}
\begin{document}
\bstctlcite{IEEEexample:BSTcontrol}

\title{robustBF: A High Accuracy and Memory Efficient 2D Bloom Filter}

\author{\IEEEauthorblockN{Sabuzima Nayak,~\IEEEmembership{Student Member,~IEEE}}
\IEEEauthorblockA{\textit{Dept. Computer Science \& Engineering} \\
\textit{National Institute of Technology Silchar}\\
Cachar-788010, Assam, India\\
sabuzima\_rs@cse.nits.ac.in}
\and 
\IEEEauthorblockN{Ripon Patgiri,~\IEEEmembership{Senior Member,~IEEE}}
\IEEEauthorblockA{\textit{Dept. Computer Science \& Engineering} \\
\textit{National Institute of Technology Silchar}\\
Cachar-788010, Assam, India\\
ripon@cse.nits.ac.in}
}

\maketitle

\begin{abstract}
Bloom Filter is an important probabilistic data structure to reduce memory consumption for membership filters. It is applied in diverse domains such as Computer Networking, Network Security and Privacy, IoT, Edge Computing, Cloud Computing, Big Data, and Biometrics. But Bloom Filter has an issue of the false positive probability. To address this issue, we propose a novel robust Bloom Filter, robustBF for short. robustBF is a 2D Bloom Filter, capable of filtering millions of data with high accuracy without compromising the performance. Our proposed system is presented in two-fold. Firstly, we modify the murmur hash function, and test all modified hash functions for improvements and select the best-modified hash function experimentally. Secondly, we embed the modified hash functions in 2D Bloom Filter.  Our experimental results show that robustBF is better than standard Bloom Filter and counting Bloom Filter in every aspect. robustBF exhibits nearly zero false positive probability with more than $10\times$ and $44\times$ lower memory consumption than standard Bloom filter and counting Bloom Filter, respectively. Source code is available at \url{https://github.com/patgiri/robustBF}.
\end{abstract}

\begin{IEEEkeywords}
Bloom Filter, Membership Filter, Approximation Algorithms, Data Structures, Algorithms.
\end{IEEEkeywords}

\section{Introduction}
Bloom Filter \cite{Bloom} is a membership filter that is capable of filtering a large number of data with a tiny amount of memory. Therefore, Bloom Filter meets diverse applications and it is applied in diverse domains, namely, Computer Networking \cite{Mun,lee}, Network Security \cite{DDoS} and Privacy \cite{PassDB1}, IoT \cite{IoT}, Big Data \cite{BigData}, Cloud Computing \cite{Singh}, Biometrics \cite{Biom} and Bioinformatics \cite{Bio}. There are diverse variants of Bloom Filter available \cite{Luo,lim,P}; however, these filters are unable to provide high accuracy using tiny memory without compromising query performance. Moreover, the false positive probability is an issue of Bloom Filter, and it cannot be reduced without increasing the memory size. There are many fast filters available, for instance, Morton Filter \cite{Morton} and XOR Filter \cite{XOR}. Morton Filter is faster than Cuckoo Filter. Morton Filter extends Cuckoo Filter and implements a compressed Cuckoo Filter. 


In this paper, we propose a novel Bloom Filter, called robustBF, is to reduce the false positive probability. robustBF is a 2D Bloom Filter, which modifies the murmur hash function for better performance. The outcomes of the proposed system are outlined as follows-
\begin{itemize}
    \item robustBF is a fast filtering Bloom Filter, and it is faster than standard Bloom Filter (SBF) \cite{Kirsch} and counting Bloom Filter (CBF) \cite{countingBF}. robustBF is $2.038\times$ and $2.48\times$ faster in insertion of 10M data than SBF and CBF respectively. 
    \item robustBF consumes 10.40$\times$ and 44.01$\times$ less memory than SBF and CBF on an average, respectively.
    \item robustBF exhibits false positive probability almost zero in the desired false positive probability setting of  0.001 with lower memory consumption. Thus, the accuracy of robustBF is almost 100\%.
\end{itemize}
We compare robustBF with SBF and CBF. We found that robustBF is better in the false positive probability and more efficient in memory consumption than the other filters' variants.  To the best of our knowledge, robustBF is the only Bloom Filter that can increase its accuracy and lower the memory footprint without compromising the filter's performance. 

The paper is organized as follows-
Section \ref{pre} discusses on the preliminaries on Bloom Filter and its terminologies. Section \ref{pro} establishes the propose systems, called robustBF. Section \ref{exp} demonstrates the performance of robustBF experimentally. Section \ref{ana} analyses on proposed systems and its memory consumption. Finally, Section \ref{con} concludes the paper.

\section{Preliminary} 
\label{pre}
Bloom Filter is introduced by Burton Howard Bloom in 1970 \cite{Bloom}. Bloom Filter has the potentiality of improving many systems. Bloom Filter is not a complete system, and it is just an enhancer of a system. Therefore, it is applied in many domains, including Networking, Security, Big Data, Bioinformatics, etc., to enhance its performance. The capability of Bloom Filter is limited to $true$ or $false$. Let, $\mathbb{B}$ be the Bloom Filter by size $m$ bits, $\mathcal{S}=\{x_1,x_2,x_3,\ldots,x_\eta\}$ be the inserted set, $\mathrm{U}$ be the universe where $\mathcal{S}\subset\mathrm{U}$ and $\eta$ be the total number of keys inserted into $\mathbb{B}$ using $\kappa$ independent hash functions. Let $x_i$ be the random query, then true positive, false positive, false negative, and true negative. There are no false negatives in the conventional Bloom Filter. The true positive, false positive, false negative and true negative are defined in Definitions \ref{def1}, \ref{def2}, \ref{def3} and \ref{def4} respectively.

\begin{definition}\label{def1}
If $x_i\in\mathbb{B}$ and $x_i\in\mathcal{S}$, then the result of Bloom Filter is called true positive.
\end{definition}
\begin{definition}\label{def2}
If $x_i\in\mathbb{B}$ and $x_i\not\in\mathcal{S}$, then the result of Bloom Filter is called false positive.
\end{definition}
\begin{definition}\label{def3}
If $x_i\not\in\mathbb{B}$ and $x_i\in\mathcal{S}$, then the result of Bloom Filter is called false negative.
\end{definition}
\begin{definition}\label{def4}
If $x_i\not\in\mathbb{B}$ and $x_i\not\in\mathcal{S}$, then the result of Bloom Filter is called true negative.
\end{definition}

\subsection{Operations}

\begin{algorithm}
\caption{Insertion of an input item $\kappa$ into Bloom Filter $\mathbb{B}$ using three hash functions. }
\begin{algorithmic}[1]
\Procedure{Insert}{$\mathbb{B}[],~\kappa,~S_1,~S_2,~S_3$}
    \State $i_1=\Call{Murmur}{\kappa,~length,~S_1}~\%~\mu$ 
    \State $i_2=\Call{Murmur}{\kappa,~length,~S_2}~\%~\mu$ 
    \State $i_3=\Call{Murmur}{\kappa,~length,~S_3}~\%~\mu$ 
    \State $\mathbb{B}[i_1]\leftarrow 1$ 
    \State $\mathbb{B}[i_2]\leftarrow 1$
    \State $\mathbb{B}[i_3]\leftarrow 1$
\EndProcedure
\end{algorithmic}
\label{Algo2.1}
\end{algorithm}

\begin{algorithm}
\caption{Lookup an item $\kappa$ in Bloom Filter using three hash functions}
\begin{algorithmic}[1]
\Procedure{Lookup}{$\mathbb{B}[],~\kappa,~S_1,~S_2,~S_3$}
    \State $i_1=\Call{Murmur}{\kappa,~length,~S_1}~\%~\mu$ 
    \State $i_2=\Call{Murmur}{\kappa,~length,~S_2}~\%~\mu$ 
    \State $i_3=\Call{Murmur}{\kappa,~length,~S_3}~\%~\mu$
    
    \If{$\mathbb{B}[i_1]~AND~\mathbb{B}[i_2]~AND~\mathbb{B}[i_3]$}
        \State $\Return~True$
    \Else
        \State $\Return~False$
    \EndIf
\EndProcedure
\end{algorithmic}
\label{Algo2.2}
\end{algorithm}

There are two key operations of Bloom Filter, namely, insert and lookup (query) operations. Algorithm \ref{Algo2.1} presents a key to Bloom Filter's insertion process using three hash function. The same process is applied in the lookup of a key, which is shown in Algorithm \ref{Algo2.2}. 

\subsection{rDBF}
\begin{definition}\label{def5}
Let, a set $\mathcal{S}=\{x_1,x_2,x_3,\ldots x_\eta\}$ and multidimensional bit array $\mathbb{B}_{d_1,d_2,d_3,\ldots,d_i}$ where $\eta$ is the total number of elements in the set $\mathcal{S}$ and $d_i$ is the dimension. The multidimensional Bloom Filter maps the elements set $\mathcal{S}$ in the multidimensional bit array $\mathbb{B}_{d_1,d_2,d_3,\ldots,d_i}$ using $k$ distinct hash functions similar to conventional Bloom Filter.
\end{definition}
r-Dimensional Bloom Filter (rDBF) is purely a multidimensional Bloom Filter that uses multidimensional Bloom array to reduce false positives and enhance computation time \cite{rDBF}. rDBF derives many variants, namely, 2DBF, 3DBF, 4DBF, 5DBF and so on, where $r=2,3,4,\ldots$ in rDBF and defined in Definition \ref{def5}. 2DBF uses two-dimensional Bloom array. Similarly, 3DBF uses three-dimensional Bloom array. rDBF uses \textbf{unsigned long int} instead of bitmap.  rDBF uses Murmur non-cryptic hash function which is the fastest hash function. Moreover, rDBF uses fast operators except hash functions and modulus operator. 

\section{robustBF- The Proposed Systems}
\label{pro}
We propose a novel Bloom Filter based on 2D Bloom Filter (2DBF) \cite{rDBF}, called robustBF. We modify the existing hash function and embed the modified hash function to constructs a new Bloom Filter. robustBF modifies the hashing technique to increase its performance. The performance of robustBF depends on the murmur hash functions, which introduces some biases and the number of hash function calls. This bias helps in improving the accuracy, and hence, reduces the false positive probability. robustBF is designed by aiming to provide high accuracy and memory-efficient Bloom Filter.  

\subsection{Selection of a hash function}
The performance of the Bloom Filter depends on the hash functions. There are many hash functions available, namely, xxHash, SuperFastHash, FastHash, FNV, and murmur hash functions. Murmur hash function is the finest hash function among all non-cryptography string hash functions \cite{Murmur}. There are also cryptographic string hash functions, namely, SHA1, SHA2, and MD5. These string hash functions are the best for security purposes. However, these cryptographic string hash functions are unable to reduce the false positive probability \cite{Eva}. Moreover, it degrades the performance of the Bloom Filter and it is proven in Patgiri \textit{et al.} \cite{Eva}. Thus, we modify the murmur hash functions and evaluated the best modification among the modified hash functions. Murmur hash function depends on the bit mixture and scanning of the input, and thus, we modify the murmur hash function and term it as H1, H2, H3, H4, H5, H6, and H7. The modified murmur hash function H1 reads 3 bytes of data at a time. Similarly, the modified murmur hash function H2 reads 4 bytes at a time. Thus, H3, H4, H5, H6, H7, H8, and H9 reads 5, 6, 7, 8, 9, 10, and 11 bytes of data at a time. It shows a huge difference in changing the reading of data at a time in murmur hash functions. The experimental results show that the H4 exhibits the best performance.  

\subsection{Designing the 2D Bloom Filter}

\begin{figure}[!ht]
    \centering
    \includegraphics[width=0.35\textwidth]{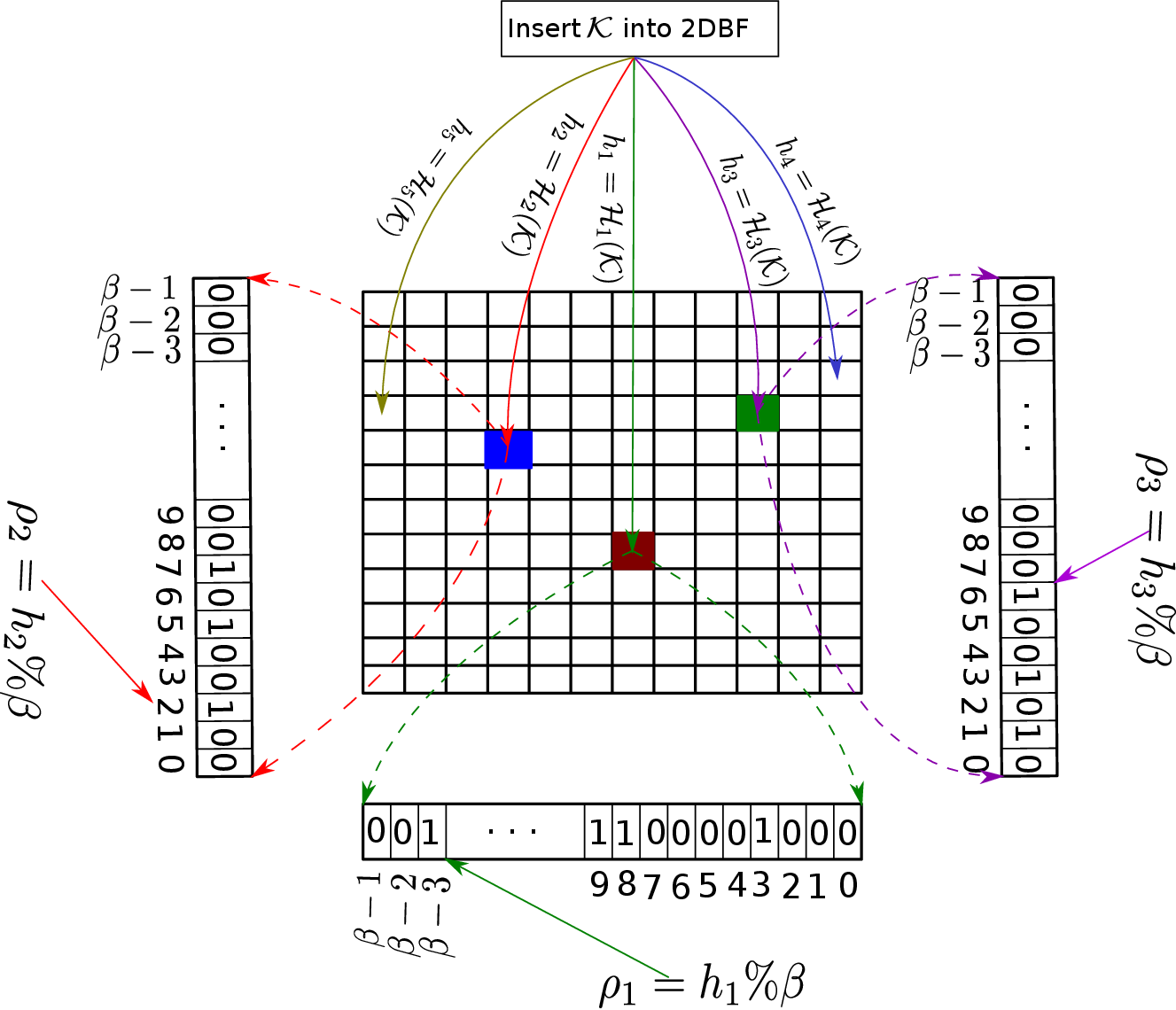}
    \caption{Insertion or lookup process in 2D Bloom Filter. }
    \label{2DBF}
\end{figure}
The detailed architecture of 2D Bloom Filter is depicted in \ref{2DBF}. Each cell of 2DBF of Figure \ref{2DBF} contains $\beta$ bits. Initially, an item is hashed into 2DBF using $k$ distinct hash functions, i.e., the item is mapped into $k$ cells of 2DBF and the bits are set to 1. Let, $\mathbb{B}_{X,Y}$ be the 2D Bloom Filter (2DBF) where $X$ and $Y$ are the dimensions, and these are prime numbers. Let, $\mathcal{H}()$ be the modified murmur hash function,  and $\mathcal{K}$ is an item to be inserted. Let, $\mathbb{B}_{i,j}$ and is the particular cell in $\mathbb{B}_{X,Y}$ filter. To insert, Equation \eqref{eq1} is invoked.
\begin{equation}\label{eq1}
    \mathbb{B}_{i,j}\leftarrow \mathbb{B}_{i,j}~OR~\mathbb{P}
\end{equation}
where $h=\mathcal{H}(\mathcal{K})$, $i=h\%X$, $j=h\%Y$, $\rho=h\%\beta$ and $\mathbb{P}=(1\ll\rho)$, where $\%$ is modulus operator, $\beta$ is the size of a cell and $\ll$ is bitwise left shift operator. The $\beta$ should be also prime number. Let us assume, \textbf{unsigned long int} occupies $64$ bits, then the the nearest prime number is $61$. Therefore, $\beta$ should be $61$ but it cannot be greater then the size of a cell in 2DBF.

The deletion operation is performed using XOR operator. To remove $\mathcal{K}$ from 2DBF, Equation \eqref{eq2} is invoked.
\begin{equation}\label{eq2}
    \mathbb{B}_{i,j}\leftarrow \left((\mathbb{B}_{i,j}~AND~\mathbb{P})==\mathbb{P}?~(\mathbb{B}_{i,j}~\oplus~\mathbb{P})~:~\mathbb{B}_{i,j}\right)
\end{equation}

For lookup operation (query operation) of an item $\mathcal{K}$, Equation \eqref{eq3} is invoked.
\begin{equation}\label{eq3}
    \digamma_{2DBF}\leftarrow (\mathbb{B}_{i,j}~AND~\mathbb{P})\gg\rho
\end{equation}
If $\digamma_{2DBF}$ is 1, then the key is a member of 2DBF and otherwise, it is not a member of 2DBF. Equation \eqref{eq1}, \eqref{eq2} and \eqref{eq3} depends on the hash function $\mathcal{H}()$. 

\section{Experimental Results}
\label{exp}
We evaluate the accuracy, false positive probability, query and insertion performance, and memory consumption of robustBF. We have conducted our experiment in Desktop PC with the configuration of Intel Core$^{TH}$ i7-7700 CPU @ 3.60GHz $\times$ 8, 8GB RAM, 1TB HDD, Ubuntu 18.04.5 LTS, and GCC- 7.5.0.


\subsection{Test cases}
We have created four different test cases to validate the accuracy, performance, memory efficiency and false positive rate. The four test cases are Same Set, Mixed Set, Disjoint Set and Random Set which are defined in Definition \ref{def6}, \ref{def7}, \ref{def8} and \ref{def9} respectively \cite{Eva}. These test cases are used to evaluate the strengths and weaknesses of Bloom Filter in every aspect. Let, $\mathcal{S}=\{s_1,s_2,s_3,\ldots,s_m\}$ be the input set and input into the robustBF. 
\begin{definition}\label{def6}
Let, $\mathcal{Q}$ is a set queried where $\mathcal{Q}=\mathcal{S}$, then the set $\mathcal{Q}$ is called Same Set.
\end{definition}

\begin{definition}\label{def7}
Let, $\mathcal{Q}=\{Q^1,Q^2\}$ be a query set where $Q^1=\{Q^1_1,~Q^1_2,Q^1_3,\ldots\}$ and $Q^1=\{Q^2_1,~Q^2_2,Q^2_3,\ldots\}$ such that $Q^1\subset\mathcal{S}$ and $Q^2\cap\mathcal{S}=\phi$, then the set $\mathcal{Q}$ is called Mixed Set.
\end{definition}

\begin{definition}\label{def8}
Let, $\mathcal{Q}$ be a query set where $\mathcal{Q}\cap\mathcal{S}=\phi$, then the set $\mathcal{Q}$ is called Random Set.
\end{definition}

\begin{definition}\label{def9}
Let, $\mathcal{Q}$ be a query set randomly generated, then the set $\mathcal{Q}$ is called disjoint set.
\end{definition}

\subsection{Experiments}
This section presents the accuracy of Bloom Filters and its memory requirements. robustBF inherits the properties from rDBF \cite{rDBF}. We compare robustBF with standard Bloom Filter (SBF) \cite{Kirsch} where SBF is a standard Bloom Filter to benchmark. The robustBF, SBF, and CBF are single-threaded Bloom Filter. Single-threaded Bloom Filters are useful as an enhancer of a system, while multi-threaded Bloom Filter is a complete system. robustBF outperforms SBF and CBF in every aspect, which is presented in this section.

\subsection{Hash function experimentation}
\pgfplotstableread[row sep=\\,col sep=&]{
interval& 10M\\
H1 & 2.84999 \\
H2 & 2.688915 \\
H3 & 2.210182 \\
H4 & 2.096695 \\
H5 & 1.989644 \\
H6 & 1.835893 \\
H7 & 1.851525 \\
H8 & 1.846735 \\
H9 & 1.845754 \\
}\looki
\begin{figure}[!ht]
\centering
\begin{tikzpicture}
    \begin{axis}[
            width=0.5\textwidth,
            height=.3\textwidth,
            enlarge x limits=0.1,
            legend style={at={(0.5,1)},
                anchor=north,legend columns=5,legend cell align=left},
            symbolic x coords={H1,H2,H3,H4,H5,H6,H7,H8,H9},
            xtick=data,
             x tick label style={rotate=45,anchor=east},
            nodes near coords align={vertical},
            ymin=1,ymax=3,
            ylabel={Time in Second},
        ]
        \addplot table[x=interval,y=10M]{\looki};
        \legend{10M}
    \end{axis}
\end{tikzpicture}
\caption{Insertion time of modified murmur hash functions H1, H2, H3, H4, H5, H6, H7, H8 and H9. Lower is better.}
\label{Hi}
\end{figure}
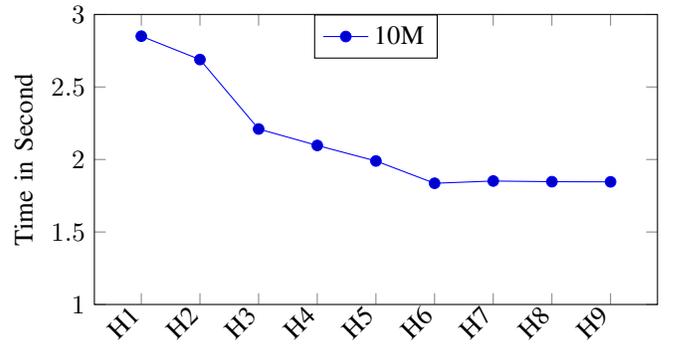

robustBF is a two-dimensional Bloom Filter and depends on the murmur hash function. The insertion performances of H1, H2, H3, H4, H5, H6, H7, H8, and H9 are demonstrated in Figure \ref{Hi}. A total of 10M data is inserted into robustBF. The performance of robustBF is the lowest in H1 and the highest in H9. However, the insertion performance is not as important as the lookup operation. 

\pgfplotstableread[row sep=\\,col sep=&]{
interval&	H1&	H2&	H3&	H4&	H5&	H6&	H7&	H8& H9	\\
Same Set&	3.300269&	2.63660&	2.19201&	2.08142&	1.980847&	1.806436&	1.811294&	1.806584&	1.810851\\
Mixed Set&	3.183355&	2.355149&	1.822252&	1.728568&	1.665192&	1.532995&	1.818304&	1.819463&	1.841217\\
Disjoint Set&	2.918455&	2.338869&	1.715064&	1.628861&	1.752703&	1.668713&	1.570839&	1.426176&	1.307276\\
Random Set&	2.767314&	2.323654&	1.799916&	1.612554&	1.722953&	1.645666&	1.537362&	1.373813&	1.267563\\
}\lookt
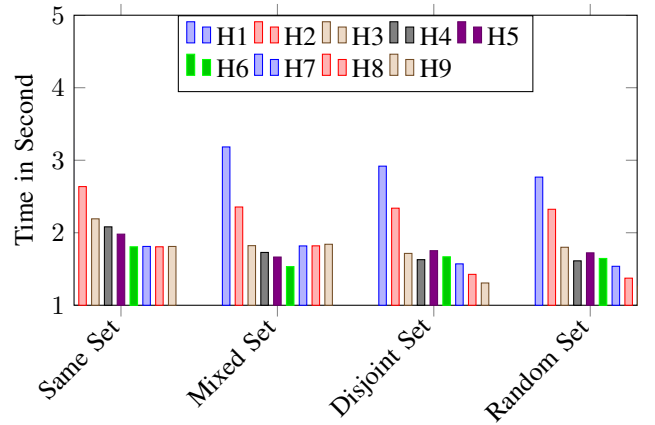
\begin{figure}[!ht]
\centering
\begin{tikzpicture}
    \begin{axis}[
            ybar,
            bar width=.1cm,
            width=0.5\textwidth,
            height=.3\textwidth,
            enlarge x limits=0.1,
            legend style={at={(0.5,1)},
                anchor=north,legend columns=5,legend cell align=left},
            symbolic x coords={Same Set,Mixed Set,Disjoint Set,Random Set},
            xtick=data,
             x tick label style={rotate=45,anchor=east},
            nodes near coords align={vertical},
            ymin=1,ymax=5,
            ylabel={Time in Second},
        ]
        \addplot table[x=interval,y=H1]{\lookt};
        \addplot table[x=interval,y=H2]{\lookt};
        \addplot table[x=interval,y=H3]{\lookt};
        \addplot table[x=interval,y=H4]{\lookt};
        \addplot table[x=interval,y=H5]{\lookt};
        \addplot table[x=interval,y=H6]{\lookt};
        \addplot table[x=interval,y=H7]{\lookt};
        \addplot table[x=interval,y=H8]{\lookt};
        \addplot table[x=interval,y=H9]{\lookt};
        \legend{H1,H2,H3,H4,H5,H6,H7,H8,H9}
    \end{axis}
\end{tikzpicture}
\caption{Lookup time of modified murmur hash functions H1, H2, H3, H4, H5, H6, H7, H8 and H9. Lower is better.}
\label{Hl}
\end{figure}

The Bloom Filter is designed to provide higher lookup performance with a tiny amount of memory usage. robustBF is evaluated its performance in Same Set, Mixed Set, Disjoint Set, and Random Set to reveal the strength and weakness. robustBF is evaluated with modified murmur hash function H1, H2, H3, H4, H5, H6, H7, H8, and H9. robustBF exhibits a similar lookup performance of 10M queries in H3, H4, H5, H6, H7, H8, and H9. Therefore, we select a modified murmur hash function from H3, H4, H5, H6, H7, H8, and H9. However, it is challenging to conclude the best performer because their performance differs in different test cases.

\pgfplotstableread[row sep=\\,col sep=&]{
interval&	H1&	H2&	H3&	H4&	H5&	H6&	H7&	H8&	H9\\
Same Set&	100&	100&	100&	100&	100&	100&	100&	100&	100\\
Mixed Set&	98.17131&	98.16534&	99.9995&	99.999&	99.99&	99.9&	50&	50&	50\\
Disjoint Set&	96.33158&	96.33909&	99.99987&	100&	99.99999&	99.99959&	95.80871&	95.80871&	95.80871\\
Random Set&	96.35987&	96.35168&	99.99981&	99.99999&	99.99994&	99.99955&	95.81099&	95.81099&	95.81099\\
}\looka
\begin{figure}[!ht]
\centering
\begin{tikzpicture}
    \begin{axis}[
            ybar,
            bar width=.1cm,
            width=0.5\textwidth,
            height=.3\textwidth,
            enlarge x limits=0.1,
            legend style={at={(0.5,1)},
                anchor=north,legend columns=5,legend cell align=left},
            symbolic x coords={Same Set,Mixed Set,Disjoint Set,Random Set},
            xtick=data,
             x tick label style={rotate=45,anchor=east},
            nodes near coords align={vertical},
            ymin=45,ymax=135,
            ylabel={Percentage},
        ]
        \addplot table[x=interval,y=H1]{\looka};
        \addplot table[x=interval,y=H2]{\looka};
        \addplot table[x=interval,y=H3]{\looka};
        \addplot table[x=interval,y=H4]{\looka};
        \addplot table[x=interval,y=H5]{\looka};
        \addplot table[x=interval,y=H6]{\looka};
        \addplot table[x=interval,y=H7]{\looka};
        \addplot table[x=interval,y=H8]{\looka};
        \addplot table[x=interval,y=H9]{\looka};
        \legend{H1,H2,H3,H4,H5,H6,H7,H8,H9}
    \end{axis}
\end{tikzpicture}
\caption{Accuracy of modified murmur hash functions H1, H2, H3, H4, H5, H6, H7, H8 and H9. Higher is better.}
\label{Ha}
\end{figure}
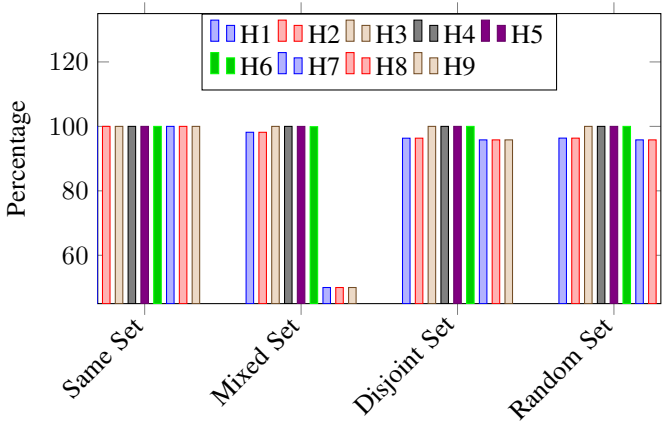

\pgfplotstableread[row sep=\\,col sep=&]{
interval&	H1&	H2&	H3&	H4&	H5&	H6&	H7&	H8&	H9\\
Same Set&	0&	0&	0&	0&	0&	0&	0&	0&	0\\
Mixed Set&	0.018287&	0.018347&	5E-06&	1E-05&	1E-04&	0.001&	0.5&	0.5&	0.5\\
Disjoint Set&	0.036684&	0.036609&	1E-06&	0&	0&	4E-06&	0.041913&	0.041913&	0.041913\\
Random Set&	0.036401&	0.036483&	2E-06&	0&	1E-06&	5E-06&	0.04189&	0.04189&	0.04189\\
}\lookfpp
\begin{figure}[!ht]
\centering
\begin{tikzpicture}
    \begin{axis}[
            ybar,
            bar width=.1cm,
            width=0.5\textwidth,
            height=.3\textwidth,
            enlarge x limits=0.1,
            legend style={at={(0.5,1)},
                anchor=south,legend columns=5,legend cell align=left},
            symbolic x coords={Same Set,Mixed Set,Disjoint Set,Random Set},
            xtick=data,
             x tick label style={rotate=45,anchor=east},
            nodes near coords align={vertical},
            ymin=0,ymax=1,
            ymode=log,
            log basis y={10},
            ylabel={False positive probability},
        ]
        \addplot table[x=interval,y=H1]{\lookfpp};
        \addplot table[x=interval,y=H2]{\lookfpp};
        \addplot table[x=interval,y=H3]{\lookfpp};
        \addplot table[x=interval,y=H4]{\lookfpp};
        \addplot table[x=interval,y=H5]{\lookfpp};
        \addplot table[x=interval,y=H6]{\lookfpp};
        \addplot table[x=interval,y=H7]{\lookfpp};
        \addplot table[x=interval,y=H8]{\lookfpp};
        \addplot table[x=interval,y=H9]{\lookfpp};
        \legend{H1,H2,H3,H4,H5,H6,H7,H8,H9}
    \end{axis}
\end{tikzpicture}
\caption{False positive probability of Lookup of modified murmur hash functions H1, H2, H3, H4, H5, H6, H7, H8 and H9. Lower is better.}
\label{Hf}
\end{figure}
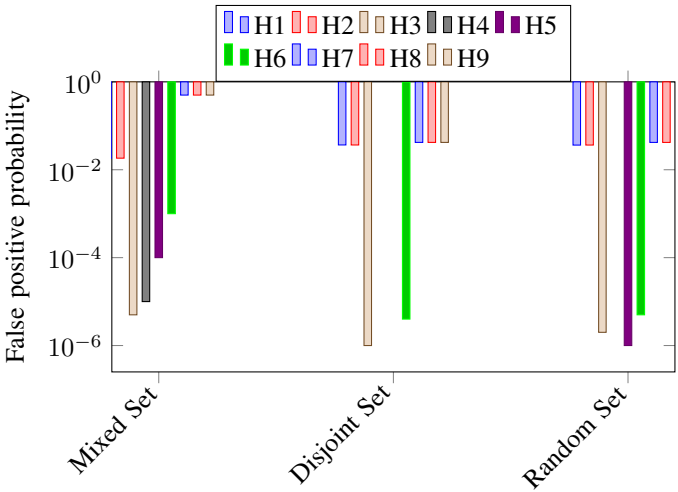

robustBF measures its accuracy as $\mathcal{A}=1-\mathcal{FPP}$ since there are no false negatives. Figure \ref{Ha} demonstrates the accuracy of robustBF using modified murmur hash functions H1, H2, H3, H4, H5, H6, H7, H8, and H9. As per our observation, the H4 has the highest accuracy and lowest false positive probability in all test cases. The false positive probability is demonstrated in Figure \ref{Hf}. The H6, H7, H8, and H9 exhibit the highest performance; however, the false positive probability of these hash functions is poor. Therefore, we are simply striking out the modified murmur hash function H6, H7, H8, and H9. The remaining hash functions are H3, H4, and H5 to compare their performance.

\pgfplotstableread[row sep=\\,col sep=&]{
interval& H5 & H6 & H7\\
10M&	2.210213&	2.096676&	1.991942\\
20M&	4.649313&	4.478097&	4.087278\\
30M&	7.167936&	6.375637&	6.192995\\
40M&	9.734465&	8.528169&	8.281784\\
50M&	12.273271&	10.738679&	10.42633\\
}\ins
\begin{figure}[!ht]
\centering
\begin{tikzpicture}
    \begin{axis}[
            width=0.5\textwidth,
            height=.3\textwidth,
            enlarge x limits=0.1,
            legend style={at={(0.5,1)},
                anchor=north,legend columns=5,legend cell align=left},
            symbolic x coords={10M,20M,30M,40M,50M},
            xtick=data,
             x tick label style={rotate=45,anchor=east},
            nodes near coords align={vertical},
            ymin=1.5,ymax=14,
            ylabel={Time in Second},
        ]
        \addplot table[x=interval,y=H5]{\ins};
        \addplot table[x=interval,y=H6]{\ins};
        \addplot table[x=interval,y=H7]{\ins};
        \legend{H3,H4,H5}
    \end{axis}
\end{tikzpicture}
\caption{Insertion time of robustBF of 10M, 20M, 30M, 40M, and 50M by H3, H4 and H5 murmur hash functions. Lower is better.}
\label{Ins}
\end{figure}
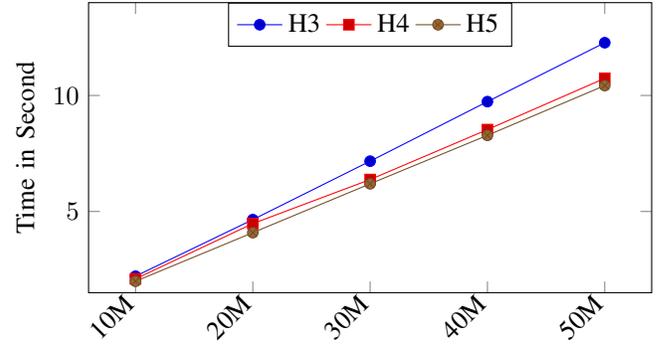

robustBF performance is measured using H3, H4, and H5 by inserting 10M, 20M, 30M, 40M, and 50M data. robustBF performance is similar to H4 and H5; however, its performance is lower in H3 as depicted in Figure \ref{Ins}. We observed that the H4 is the best performer among the modified murmur hash functions. However, H4 is not the best performer in insertion and lookup, but it is best in false positive probability. Therefore, we conclude that H4 is the best choice for robustBF. The rest of the experiment of robustBF is performed using the H4 function because H4 exhibits its best performance in the false positive probability, which is crucial for Bloom Filter.

\subsection{Comparison of robustBF with other filters}

\pgfplotstableread[row sep=\\,col sep=&]{
interval& 	robustBF&	SBF  & CBF\\
10M&	1.819094&	3.97461  & 5.199482\\
20M&	3.651776&	9.120941 & 10.78577\\
30M&	5.519504&	13.761192& 17.236443\\
40M&	7.351484&	18.236497& 23.229478\\
50M&	9.215502&	23.183817& 28.910923\\
}\ins
\begin{figure}[!ht]
\centering
\begin{tikzpicture}
    \begin{axis}[
            width=0.5\textwidth,
            height=.3\textwidth,
            enlarge x limits=0.1,
            legend style={at={(0.5,1)},
                anchor=north,legend columns=5,legend cell align=left},
            symbolic x coords={10M,20M,30M,40M,50M},
            xtick=data,
             x tick label style={rotate=45,anchor=east},
            nodes near coords align={vertical},
            ymin=1,ymax=30,
            ylabel={Time in Second},
        ]
        \addplot table[x=interval,y=robustBF]{\ins};
        \addplot table[x=interval,y=SBF]{\ins};
        \addplot table[x=interval,y=CBF]{\ins};
        \legend{robustBF,SBF,CBF}
    \end{axis}
\end{tikzpicture}
\caption{Insertion time of robustBF, SBF and CBF. Lower is better.}
\label{ins}
\end{figure}
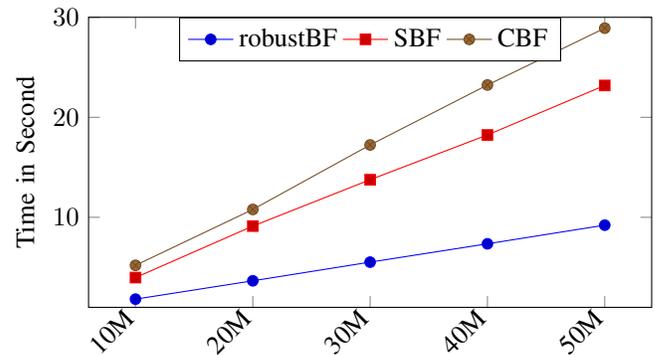

The comparison of insertion time of robustBF and SBF is demonstrated in Figure \ref{ins}. robustBF, SBF, and CBF perform 5.426 million operations per second (MOPS), 2.182 MOPS, and 1.79 MOPS on average, respectively. robustBF is faster than SBF in insertion.

\pgfplotstableread[row sep=\\,col sep=&]{
interval& 	SrobustBF&	SSBF&	MrobustBF&	MSBF& SCBF	& MCBF	\\
10M&	1.829637&	3.545045&	1.603094&	2.763179& 4.56835	& 3.513943	\\
20M&	3.625605&	7.627779&	3.108496&	6.307661& 9.583552	& 7.696776 \\
30M&	5.463271&	11.79223&	4.658303&	9.322156& 15.120955	&  11.617303\\
40M&	7.308529&	15.745596&	6.237513&	12.820064& 23.890921 & 15.857062 \\
50M&	9.183083&	19.591754&	7.782713&	15.967332& 24.531486 & 18.804615\\
}\lookup
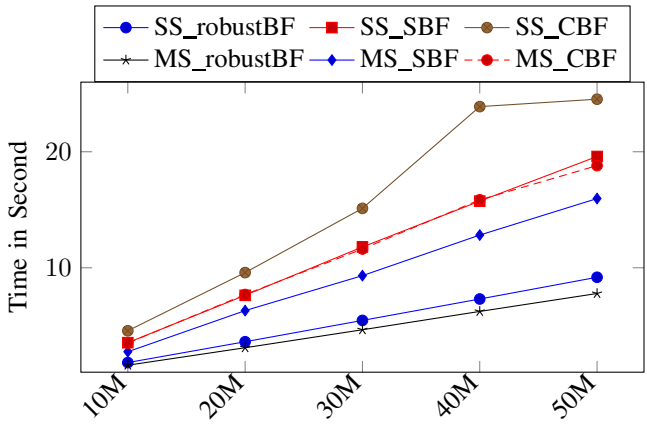
\begin{figure}[!ht]
\centering
\begin{tikzpicture}
    \begin{axis}[
            width=0.5\textwidth,
            height=.3\textwidth,
            enlarge x limits=0.1,
            legend style={at={(0.5,1)},
                anchor=south,legend columns=3,legend cell align=left},
            symbolic x coords={10M,20M,30M,40M,50M},
            xtick=data,
             x tick label style={rotate=45,anchor=east},
            nodes near coords align={vertical},
            ymin=1,ymax=26,
            ylabel={Time in Second},
        ]
        \addplot table[x=interval,y=SrobustBF]{\lookup};
        \addplot table[x=interval,y=SSBF]{\lookup};
        \addplot table[x=interval,y=SCBF]{\lookup};
        \addplot table[x=interval,y=MrobustBF]{\lookup};
        \addplot table[x=interval,y=MSBF]{\lookup};
        \addplot table[x=interval,y=MCBF]{\lookup};
        \legend{SS\_robustBF,SS\_SBF,SS\_CBF,MS\_robustBF,MS\_SBF,MS\_CBF}
    \end{axis}
\end{tikzpicture}
\caption{Lookup time of robustBF, SBF, and CBF in Same Set (SS) and Mixed Set (MS). Lower is better.}
\label{lookup1}
\end{figure}

\pgfplotstableread[row sep=\\,col sep=&]{
interval& 	DrobustBF&	DSBF&	RrobustBF&	RSBF & DCBF & RCBF \\
10M&    1.692311&	2.505276&	1.646096&	2.535557  &  2.758831 & 2.723711\\
20M&	2.944429&	5.333682&	2.84929&	5.029729  & 6.108436  & 5.913611\\
30M&	4.448056&	8.141635&	4.310109&	8.016901  &  8.956921 & 8.827946\\
40M&	6.021316&	10.743181&	5.947929&	10.542516  &  12.510179 & 12.875963\\
50M&	7.710506&	14.163228&	7.617605&	13.711169  & 15.189417  & 14.877157\\
}\lookup
\begin{figure}[!ht]
\centering
\begin{tikzpicture}
    \begin{axis}[
            ybar,
            bar width=.1cm,
            width=0.5\textwidth,
            height=.3\textwidth,
            enlarge x limits=0.1,
            legend style={at={(0.5,1)},
                anchor=south,legend columns=3,legend cell align=left},
            symbolic x coords={10M,20M,30M,40M,50M},
            xtick=data,
             x tick label style={rotate=45,anchor=east},
            nodes near coords align={vertical},
            ymin=1,ymax=16,
            ylabel={Time in Second},
        ]
        \addplot table[x=interval,y=DrobustBF]{\lookup};
        \addplot table[x=interval,y=DSBF]{\lookup};
        \addplot table[x=interval,y=DCBF]{\lookup};
        \addplot table[x=interval,y=RrobustBF]{\lookup};
        \addplot table[x=interval,y=RSBF]{\lookup};
        \addplot table[x=interval,y=RCBF]{\lookup};
        \legend{DS\_robustBF,DS\_SBF,DS\_CBF,RS\_robustBF,RS\_SBF,RS\_CBF}
    \end{axis}
\end{tikzpicture}
\caption{Lookup time of robustBF, SBF, and CBF in Disjoint Set (DS) and Random Set (RS). Lower is better.}
\label{lookup2}
\end{figure}
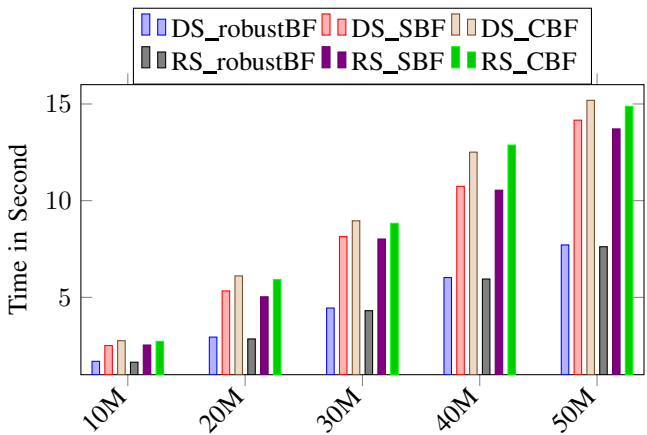

Figure \ref{lookup1} and \ref{lookup2} demonstrate the lookup time taken by robustBF and SBF in different test cases, namely, Same Set (SS), Mixed Set (MS), Disjoint Set (DS), and Random Set (RS). robustBF is faster than SBF and CBF in every aspect. robustBF performs 5.463 MOPS,	6.373 MOPS,	6.369 MOPS, and 6.423 MOPS in the Same Set, Mixed Set, Disjoint Set, and Random Set, respectively. On the other hand, SBF performs 2.554 MOPS,	3.178 MOPS, 3.667 MOPS, and 3.727 MOPS in the Same Set, Mixed Set, Disjoint Set, and Random Set on an average, respectively. Similarly, CBF performs 1.994 MOPS, 2.642 MOPS, 3.347 MOPS, and 3.384 MOPS in the Same Set, Mixed Set, Disjoint Set, and Random Set on an average, respectively.

\begin{table*}[!ht]
    \centering
    \caption{False positive probability of robustBF and SBF in various test cases. Lower is better.}\footnotesize
    \begin{tabular}{|c|c|c|c|c|c|c|c|c|c|c|}
    \hline
   Dataset&  MS\_robustBF&	MS\_SBF& MS\_CBF&	DS\_robustBF&	DS\_SBF& DS\_CBF&	RS\_robustBF&	RS\_SBF & RS\_CBF\\ \hline

10M&	0.000010&	0.001028&	0.001027&	0.000000&	0.001012&	0.000996&	0.000001&	0.000989&	0.000998\\ 
20M&	0.000010&	0.001009&	0.000996&	0.000001&	0.001002&	0.000999&	0.000000&	0.00098&	0.001002\\ 
30M&	0.000007&	0.000998&	0.000999&	0.000000&	0.001003&	0.000996&	0.000000&	0.000981&	0.000976\\ 
40M&	0.000003&	0.001019&	0.001001&	0.000000&	0.001022&	0.001003&	0.000000&	0.001&	0.000977\\ 
50M&	0.000002&	0.001013&	0.000994&	0.000000&	0.001008&	0.000999&	0.000000&	0.000976&	0.000986\\ \hline

    \end{tabular}
    \label{fpp}
\end{table*}

Table \ref{fpp} shows the false positive probability of robustBF and SBF in different test cases with 10 million to 100 million of dataset. robustBF and SBF do not exhibit any false positive probability in the Same Set. However, there is a false positive probability in other test cases. robustBF exhibits the highest false positive probability in Mixed Set, which is much lower than SBF. It is almost zero false positive probability. However, the configuration of SBF and CBF is 0.001, and hence, it shows a constantly similar false positive probability.

\begin{table*}[!ht]
    \centering
    \caption{Accuracy of robustBF, SBF, and CBF in Same Set (SS), Mixed Set (MS), Disjoint Set (DS) and Random Set (RS). Higher is better.}\footnotesize
    \begin{tabular}{|c|c|c|c|c|c|c|c|c|c|c|}
    \hline
     Interval& 	MS\_robustBF&	MS\_SBF& MS\_CBF&	DS\_robustBF&	DS\_SBF& DS\_CBF&	RS\_robustBF&	RS\_SBF & RS\_CBF\\ \hline
10M& 99.999&	99.8972&	99.8973&	100&	99.8988&	99.9004&	99.9999&	99.9011&	99.9002
\\
20M& 99.999&	99.8991&	99.9004&	99.9999&	99.8998&	99.9001&	100&	99.902&	99.8998
\\
30M& 99.9993&	99.9002&	99.9001&	100&	99.8997&	99.9004&	100&	99.9019&	99.9024
\\
40M& 99.9997&	99.8981&	99.8999&	100&	99.8978&	99.8997&	100&	99.9&	99.9023
\\
50M& 99.9998&	99.8987&	99.9006&	100&	99.8992&	99.9001&	100&	99.9024&	99.9014
\\ \hline
    \end{tabular}
    \label{acc}
\end{table*}

Table \ref{acc} demonstrates the accuracy of robustBF and SBF. Both robustBF, SBF, and CBF exhibit 100\% accuracy in the Same Set. The accuracy of robustBF is much higher than SBF and CBF. The lowest accuracy of robustBF is $99.999\%$ in 20M and 10M datasets with the Mixed Set test cases. 

\pgfplotstableread[row sep=\\,col sep=&]{
interval& 	robustBF &	SBF & CBF\\
10M&	1.558098&	17.13942 & 68.557697\\
20M&	3.232857&	34.27884 & 137.115379\\
30M&	4.992195&	51.41826 & 205.673057\\
40M&	6.59774&	68.55768 & 274.230739\\
50M&	8.251335&	85.6971 & 342.788415909\\
}\memory
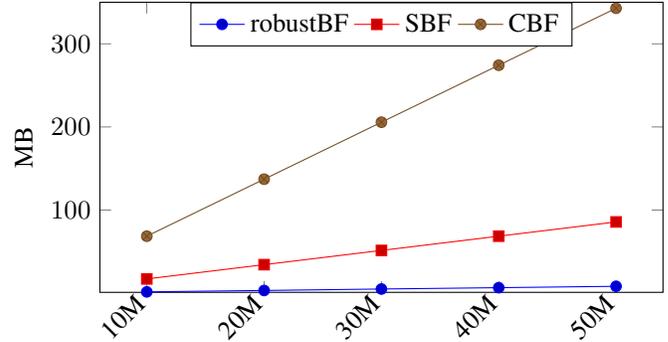
\begin{figure}[!ht]
\centering
\begin{tikzpicture}
    \begin{axis}[
            width=0.5\textwidth,
            height=.3\textwidth,
            enlarge x limits=0.1,
            legend style={at={(0.5,1)},
                anchor=north,legend columns=5,legend cell align=left},
            symbolic x coords={10M,20M,30M,40M,50M},
            xtick=data,
             x tick label style={rotate=45,anchor=east},
            nodes near coords align={vertical},
            ymin=1,ymax=350,
            ylabel={MB},
        ]
        \addplot table[x=interval,y=robustBF]{\memory};
        \addplot table[x=interval,y=SBF]{\memory};
        \addplot table[x=interval,y=CBF]{\memory};
        \legend{robustBF,SBF,CBF}
    \end{axis}
\end{tikzpicture}
\caption{Memory consumption of robustBF, SBF and CBF. Lower is better.}
\label{memory}
\end{figure}

Interestingly, robustBF uses a tiny amount of memory. SBF uses higher memory than robustBF, as shown in Figure \ref{memory}. robustBF, SBF, and CBF require 1.382 bits,	14.378 bits, and 57.511 bits memory per elements. Therefore, robustBF consumes 10.40$\times$ and 44.01$\times$ lower memory than SBF and CBF, respectively. Therefore, robustBF is better than SBF and CBF in memory consumption.

\pgfplotstableread[row sep=\\,col sep=&]{
interval&10M\\
robustBF&2.096676\\
CF&1.16536\\
}\insert
\begin{figure}[!ht]
\centering
\begin{tikzpicture}
    \begin{axis}[
            ybar,
            bar width=.1cm,
            width=0.5\textwidth,
            height=.3\textwidth,
            enlarge x limits=0.1,
            legend style={at={(0.5,1)},
                anchor=north,legend columns=5,legend cell align=left},
            symbolic x coords={robustBF,CF},
            xtick=data,
             x tick label style={rotate=45,anchor=east},
            nodes near coords align={vertical},
            ymin=1,ymax=3,
            ylabel={Seconds},
        ]
        \addplot table[x=interval,y=10M]{\insert};
        \legend{10M}
    \end{axis}
\end{tikzpicture}
\caption{Comparison of insertion time of 10M data into robustBF, and CF in second. Lower is better.}
\label{insert}
\end{figure}
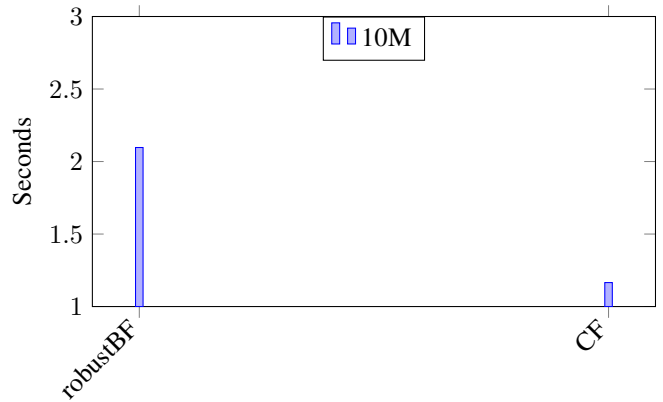

Now, let us compare robustBF with Cuckoo Filter \cite{Cuckoo} which is not a Bloom Filter. Cuckoo Filter is a membership filter based on cuckoo hashing. Figure \ref{insert} demonstrates the insertion time taken by robustBF, and CF. CF is the fastest in the insertion of 10M data than robustBF. Here, robustBF takes the highest times, which is much slower than CF.

\pgfplotstableread[row sep=\\,col sep=&]{
interval& robustBF&	CF\\
Same Set&	1.692311&	0.986211\\
Mixed Set&	1.603094&	0.98687\\
Disjoint Set&	1.692311&	1.05343\\
Random Set&	1.646096&	1.04622\\
}\lookup
\begin{figure}[!ht]
\centering
\begin{tikzpicture}
    \begin{axis}[
            width=0.5\textwidth,
            height=.3\textwidth,
            enlarge x limits=0.1,
            legend style={at={(0.5,1)},
                anchor=north,legend columns=5,legend cell align=left},
            symbolic x coords={Same Set,Mixed Set,Disjoint Set,Random Set},
            xtick=data,
             x tick label style={rotate=45,anchor=east},
            nodes near coords align={vertical},
            ymin=0.5,ymax=3,
            ylabel={Time in Second},
        ]
        \addplot table[x=interval,y=robustBF]{\lookup};
        \addplot table[x=interval,y=CF]{\lookup};
        \legend{robustBF,CF}
    \end{axis}
\end{tikzpicture}
\caption{Comparison of lookup time of 10M in robustBF, and CF in second. Lower is better.}
\label{lookup}
\end{figure}
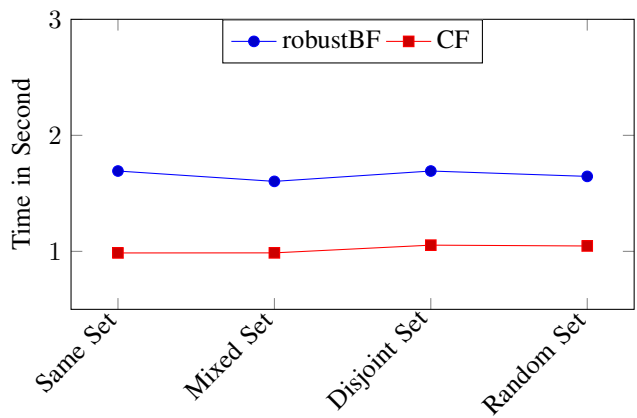

Similar to insertion, robustBF is slower in lookup than CF, as shown in Figure \ref{lookup}. CF is the fast filter to perform the lookup of 10M data.

\begin{table}[!ht]
    \centering
     \caption{Comparison of false positive probability in lookup of 10M in robustBF and CF. Lower is better.}
    \begin{tabular}{|c|c|c|c|}
    \hline
     \textbf{Test cases}& \textbf{robustBF}&\textbf{CF}\\ \hline
    Same Set&	0&	0\\ \hline
    Mixed Set&	1E-05&	0.0005592\\ \hline
    Disjoint Set&	0&	0.5749\\ \hline
    Random Set&	0&	0.995351\\ \hline
    \end{tabular}
    \label{tfpp}
\end{table}

The false positive probability of robustBF is almost zero, which is demonstrated in Table \ref{tfpp}. But, CF exhibits good false positive probability in the Same Set and Mixed set, whereas it shows the worst in Disjoint Set and Random Set of data due to the kicking process. 

\pgfplotstableread[row sep=\\,col sep=&]{
interval&10M\\
robustBF&1.558098\\
CF&24\\
}\insert
\begin{figure}[!ht]
\centering
\begin{tikzpicture}
    \begin{axis}[
            ybar,
            bar width=.1cm,
            width=0.5\textwidth,
            height=.3\textwidth,
            enlarge x limits=0.1,
            legend style={at={(0.5,1)},
                anchor=north,legend columns=5,legend cell align=left},
            symbolic x coords={robustBF,CF},
            xtick=data,
             x tick label style={rotate=45,anchor=east},
            nodes near coords align={vertical},
            ymin=1,ymax=25,
            ylabel={Seconds},
        ]
        \addplot table[x=interval,y=10M]{\insert};
        \legend{10M}
    \end{axis}
\end{tikzpicture}
\caption{Comparison of memory consumption 10M data by robustBF and CF. Lower is better.}
\label{memory1}
\end{figure}
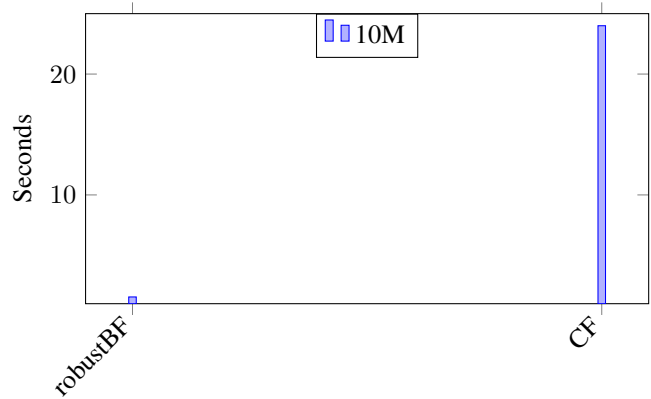

Figure \ref{memory1} demonstrates the memory consumption in 10M data by robustBF and CF. robustBF, and CF occupy 1.55MB,  and 24MB for 10M data, respectively.

\section{Analysis}
\label{ana}

Let, $m$ be the number of bits of the Bloom Filter $\mathbb{B}$ and the probability to be a particular bit is not set to 1 is $(1-\frac{1}{m})$. There are $k$ the number of hash functions and $n$ number of total items to be inserted into the Bloom Filter $\mathbb{B}$, then the probability of a bit not set to 1 is $\left(1-\frac{1}{m}\right)^{nk}$.
Therefore, the probability of a particular bit is set to 1 is
\begin{equation}\label{eq7}
    1-\left(1-\frac{1}{m}\right)^{nk} 
\end{equation}
Thus, the desired false positive probability is
\begin{equation}\label{eq8}
    \varepsilon =  \left(1-\left(1-\frac{1}{m}\right)^{nk}\right) \approx \left(1- e^{-kn/m}\right)^k
\end{equation}
The number of hash functions $k$ cannot be too large or too small. Bloom Filter requires an optimal number of hash functions. The assumption of the optimal number of hash function is $k=\frac{m}{n}ln2$.  Optimal number of hash function is an assumption and it works well with conventional Bloom Filter. It gives the optimal number of hash functions. Solving Equation \eqref{eq8} using the assumption of number of hash functions, we get
\begin{equation}\label{eq10}
    \varepsilon = \left(1- e^{-({\frac{m}{n}ln2})n/m}\right)^{\frac{m}{n}ln2}
\end{equation}
By simplifying Equation \eqref{eq10}, we get
\begin{equation}\label{eq11}
ln\varepsilon = -\frac{m}{n}(ln2)^2, ~\implies m=-\frac{n~ln\varepsilon}{(ln2)^2}
\end{equation}

The memory requirement is $m$ depending on the desired false positive probability $\varepsilon$ and number of input items $n$. robustBF uses approximately half of the desired memory calculated by Equation \eqref{eq11}. 

\subsection{Memory requirements of robustBF}
robustBF is a 2D Bloom Filter (2DBF), and thus, the memory requirement not only depends on the $n$ and $\varepsilon$ but also its dimensions of the array. Let, $X$ and $Y$ be the dimensions of robustBF. The dimensions $X\not= Y$ and these are prime numbers. If $X$ and $Y$ are not prime numbers, then false positive increases. We need to calculate $X$ and $Y$ from $m$. Let, $\mathrm{P}_i$ be the array of prime numbers and $q=\frac{m}{2*\beta}$ where the $\beta$ is the number of bits per cell in robustBF. Now, we calculate $t=\sqrt{q}$ and call the function $i=\textsc{selectPrime}(t)$ as shown in Algorithm \ref{algo1}.

\begin{algorithm}
\caption{Index calculation of prime number array $P_i$}
\begin{algorithmic}[1]
\Procedure{selectPrime}{$t$}
       \For{$i=1~to~TotalNumberOfPrime$}
            \If{$P_i>t$}
                 \Return i;
            \EndIf
        \EndFor
\EndProcedure
\end{algorithmic}
\label{algo1}
\end{algorithm}

The dimension $X=P_{i+3}$ and $Y=P_{i-3}$. Thus, robustBF allocated \textbf{unsigned long int} 2D array, i.e., the total number of memory bit is $X\times Y \times \beta$ where $\beta$ is the bit size of \textbf{unsigned long int}. Also, robustBF uses exactly half of the number of optimal hash functions calculated because it performs the modulus operation using dimensions $X$ and $Y$ to place an item in the filter.

\section{Conclusion}
\label{con}
In this paper, we present a novel Bloom Filter, called robustBF. We have demonstrated that robustBF is able to reduce the false positive probability to almost zero with desired false positive probability. Also, robustBF consumes approximately 10$\times$ and $44\times$ less memory than SBF and CBF with the same settings. We have also demonstrated its accuracy, which is satisfactorily high. The lookup and insertion performance of robustBF is higher than SBF and CBF. No Bloom filter can increase its accuracy by lowering the memory footprint significantly to the best of our knowledge. There are diverse faster filters available than robustBF. The key objectives of robustBF is to improve accuracy and lower memory footprint without compromising the filter's performance. Therefore, the robustBF performance is compared with SBF and CBF. We agree that there are many faster filter available than robustBF, for instance, Morton Filter \cite{Morton}, and XOR Filter \cite{XOR}; but these filters use large memory per items. Therefore, robustBF can improve various applications' performance using a tiny amount of memory in Computer Networking, Security and Privacy, Blockchain, IoT, Big Data, Cloud Computing, Biometrics, and Bioinformatics.

\bibliographystyle{IEEEtran}
\bibliography{mybib}

\begin{thebibliography}{10}
\providecommand{\url}[1]{#1}
\csname url@samestyle\endcsname
\providecommand{\newblock}{\relax}
\providecommand{\bibinfo}[2]{#2}
\providecommand{\BIBentrySTDinterwordspacing}{\spaceskip=0pt\relax}
\providecommand{\BIBentryALTinterwordstretchfactor}{4}
\providecommand{\BIBentryALTinterwordspacing}{\spaceskip=\fontdimen2\font plus
\BIBentryALTinterwordstretchfactor\fontdimen3\font minus
  \fontdimen4\font\relax}
\providecommand{\BIBforeignlanguage}[2]{{%
\expandafter\ifx\csname l@#1\endcsname\relax
\typeout{** WARNING: IEEEtran.bst: No hyphenation pattern has been}%
\typeout{** loaded for the language `#1'. Using the pattern for}%
\typeout{** the default language instead.}%
\else
\language=\csname l@#1\endcsname
\fi
#2}}
\providecommand{\BIBdecl}{\relax}
\BIBdecl

\bibitem{Bloom}
B.~H. Bloom, ``Space/time trade-offs in hash coding with allowable errors,''
  \emph{Comm. of the ACM}, vol.~13, no.~7, pp. 422--426, 1970.

\bibitem{Mun}
J.~H. {Mun} and H.~{Lim}, ``New approach for efficient ip address lookup using
  a bloom filter in trie-based algorithms,'' \emph{IEEE Transactions on
  Computers}, vol.~65, no.~5, pp. 1558--1565, 2016.

\bibitem{lee}
H.~{Lee} and A.~{Nakao}, ``Improving bloom filter forwarding architectures,''
  \emph{IEEE Communications Letters}, vol.~18, no.~10, pp. 1715--1718, 2014.

\bibitem{DDoS}
R.~Patgiri, S.~Nayak, and S.~K. Borgohain, ``Preventing ddos using bloom
  filter: A survey,'' \emph{EAI Endorsed Transactions on Scalable Information
  Systems}, vol.~5, no.~19, 11 2018.

\bibitem{PassDB1}
R.~Patgiri, S.~Nayak, and S.~K. Borgohain, ``Passdb: A password database with
  strict privacy protocol using 3d bloom filter,'' \emph{Information Sciences},
  vol. 539, pp. 157--176, 2020.

\bibitem{IoT}
A.~Singh, S.~Garg, S.~Batra, N.~Kumar, and J.~J. Rodrigues, ``Bloom filter
  based optimization scheme for massive data handling in iot environment,''
  \emph{Future Generation Computer Systems}, vol.~82, pp. 440 -- 449, 2018.

\bibitem{BigData}
R.~Patgiri, S.~Nayak, and S.~K. Borgohain, ``Role of bloom filter in big data
  research: {A} survey,'' \emph{International Journal of Advanced Computer
  Science and Applications(IJACSA)}, vol.~9, no.~11, 2019.

\bibitem{Singh}
A.~{Singh}, S.~{Garg}, K.~{Kaur}, S.~{Batra}, N.~{Kumar}, and K.~R. {Choo},
  ``Fuzzy-folded bloom filter-as-a-service for big data storage in the cloud,''
  \emph{IEEE Transactions on Industrial Informatics}, vol.~15, no.~4, pp.
  2338--2348, 2019.

\bibitem{Biom}
C.~{Rathgeb}, F.~{Breitinger}, H.~{Baier}, and C.~{Busch}, ``Towards bloom
  filter-based indexing of iris biometric data,'' in \emph{2015 International
  Conference on Biometrics (ICB)}, 2015, pp. 422--429.

\bibitem{Bio}
S.~{Nayak} and R.~{Patgiri}, ``A review on role of bloom filter on dna
  assembly,'' \emph{IEEE Access}, vol.~7, pp. 66\,939--66\,954, 2019.

\bibitem{Luo}
L.~{Luo}, D.~{Guo}, R.~T.~B. {Ma}, O.~{Rottenstreich}, and X.~{Luo},
  ``Optimizing bloom filter: Challenges, solutions, and comparisons,''
  \emph{IEEE Communications Surveys Tutorials}, vol.~21, no.~2, pp. 1912--1949,
  2019.

\bibitem{lim}
H.~{Lim}, J.~{Lee}, and C.~{Yim}, ``Complement bloom filter for identifying
  true positiveness of a bloom filter,'' \emph{IEEE Communications Letters},
  vol.~19, no.~11, pp. 1905--1908, 2015.

\bibitem{P}
P.~{Reviriego}, J.~{Martínez}, and S.~{Pontarelli}, ``Cfbf: Reducing the
  insertion time of cuckoo filters with an integrated bloom filter,''
  \emph{IEEE Communications Letters}, vol.~23, no.~10, pp. 1857--1861, 2019.

\bibitem{Morton}
\BIBentryALTinterwordspacing
A.~D. Breslow and N.~S. Jayasena, ``Morton filters: Faster, space-efficient
  cuckoo filters via biasing, compression, and decoupled logical sparsity,''
  \emph{Proc. VLDB Endow.}, vol.~11, no.~9, p. 1041–1055, May 2018. [Online].
  Available: \url{https://doi.org/10.14778/3213880.3213884}
\BIBentrySTDinterwordspacing

\bibitem{XOR}
T.~M. Graf and D.~Lemire, ``Xor filters: Faster and smaller than bloom and
  cuckoo filters,'' \emph{ACM J. Exp. Algorithmics}, vol.~25, Mar. 2020.

\bibitem{Kirsch}
A.~Kirsch and M.~Mitzenmacher, ``Less hashing, same performance: Building a
  better bloom filter,'' \emph{Random Struct. Algorithms}, vol.~33, no.~2, p.
  187–218, Sep. 2008.

\bibitem{countingBF}
L.~Fan, P.~Cao, J.~Almeida, and A.~Z. Broder, ``Summary cache: A scalable
  wide-area web cache sharing protocol,'' \emph{IEEE/ACM Trans. Netw.}, vol.~8,
  no.~3, pp. 281--293, Jun. 2000.

\bibitem{rDBF}
R.~Patgiri, S.~Nayak, and S.~K. Borgohain, ``{rDBF}: A r-dimensional bloom
  filter for massive scale membership query,'' \emph{Journal of Network and
  Computer Applications}, vol. 136, pp. 100 -- 113, 2019.

\bibitem{Murmur}
A.~Appleby, ``Murmurhash,'' Retrieved on September 2020 from
  https://sites.google.com/site/murmurhash/, 2020.

\bibitem{Eva}
R.~{Patgiri}, S.~{Nayak}, and N.~B. {Muppalaneni}, ``Is bloom filter a bad
  choice for security and privacy?'' in \emph{2021 International Conference on
  Information Networking (ICOIN)}, 2021, pp. 648--653.

\bibitem{Cuckoo}
\BIBentryALTinterwordspacing
B.~Fan, D.~G. Andersen, M.~Kaminsky, and M.~D. Mitzenmacher, ``Cuckoo filter:
  Practically better than bloom,'' in \emph{Proceedings of the 10th ACM
  International on Conference on Emerging Networking Experiments and
  Technologies}, ser. CoNEXT '14.\hskip 1em plus 0.5em minus 0.4em\relax New
  York, NY, USA: Association for Computing Machinery, 2014, p. 75–88.
  [Online]. Available: \url{https://doi.org/10.1145/2674005.2674994}
\BIBentrySTDinterwordspacing

\end{thebibliography}
\end{document}